\documentstyle{article}
\begin{document}

\begin{center}
         {\Large
				 Novel Cholesteric Phase \vspace{5mm}  \\
				 in Dispersions of  Nucleic Acids   \vspace{5mm}  \\
												 due to  \vspace{5mm}	\\
                                                         Polymeric Chelate      Bridges
	 }
\vspace{2cm} \\
							 {\large V.L. Golo}$^{\dagger}$,
							 {\large E.I. Kats}$^{\dagger \dagger}$,
							 {\large Yu.S. Volkov}$^{\dagger}$, 
                                     {\large V.I. Salyanov}$^{\dagger \dagger \dagger}$
 \vspace{3mm}  \\
                                      and
 \vspace{3mm} \\
							 {\large Yu.M. Yevdokimov}$^{\dagger \dagger \dagger}$
							 \vspace{1cm}\\
							 $^{\dagger}$ Department of Mechanics and Mathematics \\
													Moscow University, Moscow 119899, Russia \vspace{5mm} \\
							 $^{\dagger \dagger}$ Landau Institute for Theoretical Physics  \\
													 Russian Academy of Sciences    \\
													 Moscow, Russia \\
                                                             and Institute Laue---Langevin,  \\ 
                                                             BP --- 156, Grenoble, Cedex --- 9, France \vspace{5mm} \\
							 $^{\dagger \dagger \dagger}$ Institute of
													 Molecular Biology \\
													 Russian Academy of Sciences \\
											 Moscow, Russia
\end{center}

\noindent
{\Large  Abstract } \vspace{5mm}

We consider cholesteric liquid-crystalline  DNA dispersions, and show
that polymeric (Dau-Cu) complexes, the so-called bridges,  between pairs of
DNA molecules may generate a super liquid-crystalline structure (BR-phase).
The latter could have a layered spatial structure and an abnormal optical
activity that could have a bearing upon the intense CD-band observed
in DNA-dispersions.

\pagebreak

\noindent
{\bf 1. Introduction} \vspace{1cm}

\noindent
The existence  of the liquid-crystalline (LC)-phases of DNA
molecules complexed
with antibiotics of the anthracycline family opened a new avenue to
studying both DNA and solvents.
In this respect circular dichroism (CD)-experiments using
daunomycin (Dau) are very interesting, \cite{dau1}, \cite{dau2},
\cite{dau3}, owing to the (Dau-DNA) complexes having very specific
structural properties and optical activity.
It is alleged,\cite{BCMS}, \cite{FrB}, that Dau-molecules  may intercalate
between base pairs of the double-stranded DNA fixed in liquid-crystalline dispersion, 
forming the so-called
 intercalation  complex (I),  
and generate an  additional set of chromophores
contained in a DNA-molecule, that is besides the original ones
due to nitrogen bases absorbing in the UV region of $ 270 \;~ nm$,
there emerge chromophores, belonging to  Dau-molecules,
that absorb in the region of $500 \;~ nm$, see \cite{BCMS}, \cite{BBHJ},   \cite{PFH}     
for the detailed analysis of 
optical phenomenon involved.
As the concentration of daunomycin increases,
the intensity of the CD-band at $500\;~ nm$ rises, until it reaches
a saturation point.  According to \cite{FrB} the saturation of the
CD-intensity correponds to the ratio of one Dau molecule per five base
pairs, see Fig.1. Further adding the daunomycin results only in a
larger number of Dau molecules that are not attached to DNA molecules.
These free Dau molecules do not produce a coherent effect as regards the
CD absorption, and therefore do not have a bearing upon the CD intensity,
in agreement with the saturation effect mentioned above.
It is worth noting that one can change the sign of both CD bands, \cite{bustamante}, at
$270\; nm$ and $500\; nm$ by increasing the concentration of Dau
above a certain point.  Since the sign of CD band corresponds to the
sense of the helical twist, i.e. the left/right one
for the negative/positive
band, respectively, one may conclude that Dau molecules
generate a structural transition
 of the cholesteric DNA phase.

The conformation of (DNA-Dau) complex is substantially modified,
if the dispersion is treated with $CuCl_2$, and therefore
copper ions, $Cu^{2+}$, are added in the system.
Then according to the X-ray analysis, \cite{FrB}, the Dau-saturation
discussed above having been reached, there appear Dau-molecules,
{\it adsorbed} at the surfaces of the DNA-double helices fixed in the cholesteric phase,
the so-called external complex II, see Fig. 2 . 
If the concentration of
 $Cu^{2+}$ is large enough, 
they result in the formation of chelate complexes. 
In case the chelate complexes are properly located with respect to the double helices
of adjacent DNA molecules, there arise certain planar structures, or bridges, that
cross-link pairs of adjacent
 DNA-molecules, \cite{FrB}, see Figs. 1 --- 3, \cite{GD} --- \cite{ML}. 
It is generally accepted
that the bridges have the form of polymeric chelate complexes
\begin{equation}
	\left [ DNA \, - \, Dau \, - \, (\, Cu \, - \, Dau \, - \, Cu \,)_n
					\, - \, Dau \, - \, DNA \right ] \label{chbridge}
\end{equation}

The amplitude of CD-bands increases tens times after
the Dau- and the copper concentrations
having reached threshold values, \cite{dau2}, see Fig.4.
Here it should be noted that for one thing
isolated Dau-molecules, as well as other isotropic aggregates, are not
sufficient to produce an abnormal optical acitvity, and for another
there are relatively few Dau-molecules that are properly oriented so as to
form chelate bridges.
Thus, even though
according to the arguments given above, the thresholds could correspond to
the formation of the chelate bridges, the total effect due to Dau-molecules
and their chelate complexes considered separately could be too small.
Therefore, it is reasonable to
suggest that the observed increase in the CD-bands be due to a cooperative
effect related to chelate bridges, and  one should look for it in
their spatial organization.  Even though the anisotropy of the chelate
bridges is small, they could acquire an orientational ordering while
serving as a kind of linker between DNA-molecules. Thus, we may suggest
that cholesteric dispersions of DNA doped with $Cu^{2+}$ could be
visualized as two LC systems in interaction; the DNA molecules forming
the dominant part of them.   \vspace{2cm}

\noindent
{\bf 2. Double Cholesterical Phase} \vspace{1cm}

\noindent
In what follows we shall not take into account the influence of ambient solvent on 
DNA molecules and polymeric chelate bridges.
We may cast the arguments given in the Introduction in a more quantitative form by employing the
familiar macroscopic formalism of the order parameter and
assign  a unit vector $ \vec n $ to the cholesteric phase formed by
the DNA-molecules, and a unit vector $ \vec \nu $ to the bridges.
Initially, we do not make any assumption about their forming a nematic or
cholesteric phase; their orientational ordering is to a large extent
determined by the molecules of DNA.
Therefore, we assume that the free energy of the compound system
formed by DNA and bridges, should contain an interaction term given by
the following equation
\begin{equation}
 F_I = \int d^3 \vec r \, \lambda\,
				\left ( \vec \nu \cdot rot\, \vec n \right )^2         \label{frn}
\end{equation}
in which the real unit vectors $\vec n$, $\vec \nu$ are the order
parameters for the DNA and the bridges, respectively. It is important that
the expression given above is the simplest one for the interaction energy
that takes into account the cholesteric twist of the phase due to
DNA-molecules, and its possible interaction with bridges.
Next, we may write down the part of the free energy related
to the DNA-molecules considered as a separate system, and  since we aim at
deriving only a qualitative model, we may employ one dimensional
approximation and write
\begin{equation}
	F_{DNA} = \int \frac{k}{2}\, \left ( \vec n \cdot rot\, \vec n \: - \: q_0
															\right ) ^2 \, dz\; ,  \label{dna}
\end{equation}
that is the usual form accommodating cholesteric twist.

Again employing the most simple terms,  we assume that
the bridges have only the orientational energy, proper, the cholesteric
effects being  due to their interaction with the
DNA-molecules; therefore, we write
\begin{equation}
	F_{Br} = \int \frac{\mu}{2}\, \left (  \frac{d}{dz}\, \vec \nu
									 \right ) ^2 \, dz  \label{br}
\end{equation}

On combining equations (\ref{frn}), (\ref{dna}), and (\ref{br}),
we obtain the equation for the free energy of the whole system

\begin{equation}
		F = \int \left [ \frac{k}{2}\, \left( \vec n \cdot rot\, \vec n\,
																-\, q_0	 \right ) ^2 \, +
																	 \frac{\mu}{2}\,
																	 \left( \frac{d}{dz}\, \vec \nu  \right)^2\,
																	 +\, \lambda\,
																	 \left ( \vec \nu \cdot rot\, \vec n
																	 \right ) ^2\,
							\right ]\,  dz   \label{F}
\end{equation}

\noindent
in which the z-axis is assumed to be the helical one of the DNA-cholesteric,
so that the vector $\vec n$ can be cast in the form
\begin{equation}
	 \vec n = (\cos \phi, \sin \phi, 0) \label{phi}
\end{equation}
in which the polar angle $\phi$ is a function of $z$ and
$$\frac{d \phi}{dz}$$
is the twist angle. The vector $\vec \nu$ reads
\begin{equation}
		\vec \nu = (\sin \theta\, \cos \psi,\:
		\sin \theta\, \sin \psi,\: \cos \theta ) \label{psi}
\end{equation}

The angle $\theta$ is, to a certain extent, a joint characteristic of
the position of adjacent DNA-molecules and chelate bridges linking them,
In fact, let us consider two molecules of DNA which we may visualize as
two skew straight lines, $\Lambda_1$ and $\Lambda_2$, at a distance $d$
from each other;
$d$ being  equal to the distance between two parallel planes, $\Pi_{1,2}$,
that contain the lines $\Lambda_{1,2}$, respectively, see Fig.5.
If the length of a bridge, $l$, and the distance $d$ are fixed,
the angle $\theta$ is fixed as well. In real life, there are fluctuations
of the positions of the planes  $\Pi_{1,2}$ that produce a change in $d$.
Also, the length of bridges may vary, for the structural form given by
Eq.(\ref{chbridge}) may correspond to different values of the number
of $Cu$-units, $n$.
According to X-ray analysis \cite{dau2}, in the absence of the bridges,
the change in $d$  amounts to 1\%.
According to a crude and qualitative nature of our
model we shall consider only sufficiently small fluctuations of $d$,
and assume the angle $\theta$ fixed.

Thus, let us assume that $ \theta = \theta_0 = const $.
Then the equation for the free energy (\ref{F}) acquires the form
\begin{equation}
	F = \int \left [ \frac{k}{2}\,  (\dot \phi - q_0)^2\,
						+\, \frac{\mu}{2}\, \sin^2 \theta_0\, \dot \psi^2\,
						+\, \lambda\, \sin^2 \theta_0\, \cos^2(\phi - \psi)\,
																						\dot \phi^2
					 \right ]\, dz  \label{Fangles}
\end{equation}
\noindent
in which the dot means the differentiation with respect to $z$. The
minimization of equations for the free energy given by Eq.(\ref{Fangles}) reads
\begin{eqnarray}
	\ddot \phi & = &  \frac{\lambda \sin^2 \theta_0\, \sin[2(\phi - \psi)]}
												 {k + 2 \lambda \sin^2 \theta_0\, \cos^2(\phi - \psi)}\;
												 (2 \dot \phi \dot \psi - \dot \phi^2)  \label{main} \\
	\ddot \psi & = &  \frac{\lambda}{\mu}\; \sin[2(\phi - \psi)]\; \dot \phi^2     \nonumber
\end{eqnarray}
If there are no bridges, i.e. $ \mu = \lambda = 0 $, we have the
usual solution $ \dot \phi = q_0 $ corresponding to the minimum of $F$ and
giving the cholesteric structure.
We assume that the terms related to the bridges are small in comparison
with the main cholesteric term
$$
	 \frac{k}{2}\, ( \dot \phi -  q_0 )^2\; ,
$$
and measure the constants
$\nu$ and $\lambda$  in units of $k$; the latter being of order
$10^{-6} \, dyn$.  There is no experimental information about the values of
$\mu$ and $\lambda$; our guess is that for one thing $\mu$ is much smaller
than $k$, $\mu \ll k$, owing to a small anisotropy of the molecules of
bridges, at least in comparison with those of DNA,
and for another the
magnitude of $\lambda$  approaches that of $k$, even though being smaller,
for the interaction term should be strong enough to drive the bridges along
the orientational twist of DNA molecules.
In fact, one can visualize
a chelate bridge as a planar rectangle with the ratio of the sides
about 3 by 10 \AA, so that its anisotropy is much smaller than
that of DNA-molecules.
Therefore, we shall take $k=1$
and $\mu$ of the order of $0.01$ and $\lambda$ of $0.1$,
in the units given above.

As regards functional (\ref{Fangles}),
we do not have the ambition to look for its minimum,
but even studying the necessary condition for the latter,
provides valuable information about the structure of the system.
In fact, there is an exact solution to the minimization equations for
(\ref{Fangles}), but its formulas are not particularly tractable, and
it is more practical to employ computer, from the very beginning.
For our purpose we need to consider
the behaviour of the twist angles, that is $\dot \phi$ for the DNA-molecules,
and $\dot \psi$ for the bridges. To understand their possible configuration,
let us visualize two adjacent molecules of DNA as two skew straight lines,
see Fig.5. It is easy to convince oneself that the  locus of points for
the middle of the segment AB of a given length $l$ is a rather elongated
ellipse $\cal E$, see Figs. 6 and 7.

It is worth noting that the ellipse is "flat" proportionally to the
smallness of the twist angle $\Phi$ between the skew lines $\Lambda_{1,2}$.
The z-axis being the rotation axis of the helical structure for the
DNA-cholesteric, the straight lines $\Lambda_{1,2}$ rotate about $Oz$
at the angular velocity $\dot \phi$, while an observer moves along $Oz$.
In the meantime the segment AB,
corresponding to a bridge, participates in two motions; that of the
straight lines, and a motion of its own along the
ellipse $\cal E$. The latter has its own structure as to the sign of
the motion along $\cal E$ and characteristic time, or period.
Of course, we do not mean the motion in time, our problem is the statical one,
and there is no time, but only a change in the charateristic of the system
as we move along the helical axis, Oz, which is the rotation axis of the
cholesterical structure. It is important that the spatial conformation of
the bridges and the DNA-molecules is determined by equilibrium conditions
imposed on the system, and is of thermodynamical nature.

Thus, our hypothesis is that the orientational ordering of the bridges
could substantially differ from that of the DNA-molecules.
The numerical simulation of Eqs.(\ref{main}) supports the claim.

The behaviour of the twist angles $\dot \phi$ and $\dot \psi$ is
illustrated in Fig. 8. We see that they can differ in magnitude
several times, and in sign as well. We can visualize the phenomenon with
the help of Figs.6 and 7 for the motion of the segement AB, or bridge, round
the supporting ellipse $\cal E$. In fact, the comparison of the numerical
simulation given in Figs. 8, and Figs.4,5 indicates that the periods
during which the twists $\dot \phi$ and $\dot \psi$ are almost identical
correspond to the location of the segment AB, or bridge, at the aphelion and
the perihelion of the ellipse $\cal E$. On the contrary, the flat regions of
the ellipse correspond to the motion of the bridge at its own angular
velocity, which is several times larger than that of the twist $\dot \phi$.
It should be noted that we do not mean an actual motion in time, but
only the change in variables as we move along the Oz-axis.

Thus, we may expect that in a dispersion of DNA cross-linked by
chelate bridges there is a super helical structure due to the bridges,
which is in a way superimposed on the cholesteric one of DNA molecules, see
Fig. 9.  In fact, the numerical simulation shows that there is a layered
orientational structure of bridges that has the property of a cholesteric
with a variable twist angle. It admits of  the abnormal
CD-band discussed in the Introduction. In fact, it is alleged that
the absorption strongly depends on the twist angle of the cholesteric structure,
and therefore, the cholesteric twist for bridges being
several times larger than that for DNA, one may expect that the CD-band
could be larger than for the cholesteric phase of DNA
in the absence of them. \vspace{2cm}

\pagebreak

\noindent
{\bf 3. Conclusion: cholesterical BR-phase} \vspace{1cm}

\noindent
We suggest that in dispersions of DNA containing chelate bridges
there could exist, besides the cholesterical phase of
DNA molecules, another cholesterical phase generated by the bridges,
which we shall call the BR-phase, see Fig.9.
The latter could be the cause for
the abnormal optical activity observed in CD-spectra in the region of $500\; nm$, and
should have a layered structure.

It should be noted that, as follows from our crude model, the cholesteric pitch
of the DNA dispersion does not change under the influence of the bridges,
at least for the values of parameters considered in this paper.
The circumstance has an important bearing upon the optical properties of
DNA dispersions.  In fact, the intensity of the CD band at $270\; nm$,
which corresponds to DNA chromophores, should have remained unaltered in the
presence of the bridges, but it increased.   Therefore, we  shall conclude
that the intensity of the CD bands depends also on certain factors besides
the orientational ordering of the bridges.  Perhaps, the chelate bridges
could change the coupling of oscillating dipole moments of adjacent DNA
molecules so that collective exciation modes
should emerge and result in the observed large intensity of the CD bands.

In this respect it is worth noting
that the use of ions of different transition metals, e.g.
cadmium or nikel, instead of copper,  results in different properties
of the BR-phase, as regards, at least, its optical activity.
The phenomenon could have an important
bearing upon the study of DNA-dispersions and serve as a kind of probe into
the physics  of DNA and solvents. For example, it would be extremely
interesting to find compounds capable of changing the sign of
CD-absorption band; the picture given above  of a chelate bridge
"moving"  along the supporting ellipse in the opposite direction with
respect to the cholesteric twist of DNA, suggests that it could be possible.

\pagebreak

\end{document}